\documentstyle[12pt,ssi,times]{article}

\begin{document}

\parskip 0.15truein

\title{The role of supersymmetry phenomenology in particle physics}

\author{James D. Wells\thanks{
Supported in part by the Department of Energy and the Alfred P. Sloan
Foundation
\vskip 0.5in 
\noindent
\copyright\ 2000 by James D. Wells. }\\ 
Physics Department, University of California, Davis, CA 95616 \\
Theory Group, Lawrence Berkeley National Lab, Berkeley, CA 94720 
\\[0.4cm]
}

\maketitle
\begin{abstract}%
\baselineskip 16pt 

Supersymmetry phenomenology is an important component of particle
physics today.  I provide a definition of supersymmetry phenomenology,
outline the scope of its activity, and argue its legitimacy.  This essay
derives from a presentation given at the
2000 SLAC Summer Institute.

\end{abstract}


\section{Introduction}

The talk I delivered on ``Supersymmetry Phenomenology'' 
at the 2000 SLAC Summer Institute was a 
relatively standard talk on the definitions, motivations
and research activities associated with supersymmetry phenomenology.
Being the energetic conference that SSI is, much of the interesting 
discussion occurred after my talk.  Participants
asked me many challenging questions about supersymmetry and supersymmetry
phenomenology. I enjoyed these exchanges and think they were
at least as useful to me and the participants as was the talk.  
In this proceedings write-up, I would therefore like to emphasize
the relevant post-talk issues as an addendum to the
actual talk~\cite{Wells talk}.

The purpose of this write-up is not to restate the meaning of supersymmetry
and the basic moving parts of the theory.  Some of this was covered
in the talk, and much of it is covered admirably in pedagogical
reviews of the subject~\cite{reviews}.  Instead, I wish to enter
quickly and directly into my view of supersymmetry phenomenology.

We should start with a definition of ``supersymmetry phenomenology''
that is
accurate, useful, and inclusive for all those who view themselves as 
working on the subject. 
{\bf Supersymmetry phenomenology attempts to answer four questions:
(1) How can supersymmetry 
account for phenomena already measured and quantified? 
(2) How can supersymmetry resolve 
its own induced problems? (3) How can 
supersymmetry be found at future experiments?  And, 
(4) how can supersymmetry be killed?}

If we look at any one of those four elements of supersymmetry phenomenology
in isolation, the field appears unmoored and speculation chasing.  For example,
if the field were interested only in question {\bf (3)}, the whole study
would be without motivation.  If the field were only concerned
with question {\bf (1)} practitioners would be charged with engaging in
post-facto natural philosophy rather than scientific inquiry.  The legitimacy
of supersymmetry phenomenology
derives from its commitment to address {\it all}
questions raised above, and, just as important, 
the studies to date demonstrate very encouraging answers to those questions.

I would like to discuss, essay style, each of the questions in turn, and
try to convince and remind the reader how much progress has been
made answering them, and how much more work is still required.

\section{How can supersymmetry account for the known?}

I would not say any known phenomena begs for supersymmetry, but a
supersymmetric theory appears to make the natural
world more understandable.  For example, we
know that the Higgs sector in the standard model has quadratic
divergences and is likely unstable to non-trivial physics
at high mass scales.  Supersymmetry elegantly solves this instability
problem by protecting hierarchies with a symmetry~\cite{reviews}.

Furthermore, the renormalization group evolution of Higgs masses
in supersymmetry naturally induces electroweak symmetry breaking.
This is often called ``radiative electroweak symmetry breaking'' because
the radiative corrections (RGE evolution) propel the Higgs mass squared 
to negative values.  And most interestingly, the minimal supersymmetric
standard model, with a large top Yukawa coupling, generically predicts
only one scalar mass squared going negative (up-Higgs), whereas
all the remaining scalar masses (Squarks, sleptons, sneutrinos) remain
positive. One can almost raise electroweak symmetry breaking to the level
of a prediction of an $SU(3)\times SU(2)\times U(1)$ supersymmetric
theory with our known particle content and Yukawa couplings.  This is
perhaps too strong, but one should realize that radiative electroweak symmetry
breaking was discovered and understood~\cite{radiative} 
well after supersymmetry was 
introduced as a possible component to nature.  As we will see in several
other examples, this pattern of finding good things flowing from a 
supersymmetry hypothesis is encouraging.

Another example of supersymmetry's success is the explanation for
the ratios of gauge couplings and Yukawa couplings we measure.
Gauge coupling unification and Yukawa unifications are based on
model dependent assumptions extending beyond the simplest supersymmetrizing
of the standard model.  Namely, they imply unification --- either Grand
Unifications (GUTs) or some type of string unification. Yukawa unification,
and the predictions derived from that are highly sensitive to matter
content at the high scale.  For this reason I will not attempt to force
my positive opinions about this on others.  However, I think the
success of gauge coupling unification is extraordinary and wish to
make a few comments on this.

Most of us have seen ad nauseum the graphs of the three gauge couplings
unifying at a high scale in supersymmetry but not in the standard model.
I think this suggests supersymmetry is the right way to go.
Of course, it could be an accident that all three gauge couplings crossed
at the same place in supersymmetry. Let's for a moment assume that there
is nothing significant about that.  Observers must also witness
another accident: the three gauge coupling RGE trajectories
cross at a scale that is not too low to be ruled out by proton decay 
experiments {\it and} cross
not too high to be meaningless when combined with gravity.  To me, this 
constitutes another interesting accident.  Just like on the roadways,
one accident is a curiosity, two accidents side-by-side is a real effect 
requiring deeper understanding of what happened.

It is true that the more accurate measurements of the last few years at LEP
and SLC clearly show that extrapolation to high scales gives a small
mismatch of couplings at the unification scale in a supersymmetric theory.  
That does not mean
there is a problem with gauge coupling unification.  It is almost certain
that there are threshold corrections at the high-scale that will
not enable one to simply take low-scale couplings and get exact unification
at the high scale. There {\it must} be a slight mismatch.  However, if
the unification is perturbative, and the GUT group does not contain
extraordinary amounts of matter, the mismatch should only be a few percent.
Indeed, this is what is found.  

The success of gauge coupling unification in a particular theory is measured
by the three gauge couplings being ``within specs'' (within the variations
inducible by threshold corrections) at the high scale.  Supersymmetry
easily passes this test, theories springing from the standard model do not.

\section{\bf How can supersymmetry resolve its own problems?}

As with many things in life, an apparent solution to one problem can
lead to even more problems.   Buying a dog to protect you from burglars
may only lead to your couch being chewed up.
Whenever you introduce a new theory (supersymmetry) to supplant
an old theory (standard model),
you run the risk of spoiling all the successful predictions and
explanations that we've held dear for so long.  For example, small
flavor changing neutral currents (FCNC)
seen by experiments are mostly natural within the standard model.  
Supersymmetric descriptions, on the surface, have no preference for
small FCNC effects over large effects.
This is a potential problem.

However, the resolution to many of these questions is wrapped up in 
supersymmetry
breaking.  Some supersymmetry breaking mechanisms, such as gauge mediated
supersymmetry breaking, automatically solve many of these types of 
problems~\cite{giudice}.
Some theories do not automatically, or naturally, solve the problems.
What we must keep in mind is that supersymmetry is not one theory --- 
it's a multitude of related theories (with only some
non-negotiable properties in common) that each have strengths and weaknesses
in describing what we already know.  The supersymmetric theories that
only have weaknesses get ignored over time.

As one illustration of how supersymmetry phenomenology studies have better
understood how internal inconsistencies can be resolve, we turn to
proton stability.  In the standard model, lepton number conservation
and baryon number conservation is an accidental global symmetry at
the renormalizable level.  For this reason, the proton is remarkably
stable within the standard model framework.

In supersymmetry, as is the case in
most beyond-the-standard model theories, proton stability is not even
approximately automatic.  The most general gauge invariant, Lorentz invariant
lagrangian does not respect baryon and lepton number conservation.  Somehow
they must be banished.
We all have a sense
of what it means to solve a problem, and part of that solution is not to
add epicycles to
a theory that is apparently breaking down.  
However, supersymmetry phenomenologists have identified several
elegant explanations for proton stability.  One such explanation is
R-parity.

R-parity is well-motivated in
that it is a simple $Z_2$ symmetry that has no discrete gauge anomalies,
and so could be derivable from a gauge symmetry~\cite{discrete}.  
This is important for
our confidence that the discrete symmetry has a solid underpinning, and won't
be violated by quantum gravity effects.  Furthermore, the requisite 
continuous gauge
symmetry is $U(1)_{B-L}$ which is contained in many of our formulations
of grand unified theories.  We can see R-parity as coexisting nicely
with grand unified theories, gauge coupling unification, and proton stability,
all within the same theoretical framework.
For these reasons, R-parity is not an ad hoc assumption in my view.

The introduction of R-parity has another important, non-trivial implication:
the lightest supersymmetric particle is stable.
Of course, stable particles can be cosmological
relics, and the first discussion of stable LSP relics 
exclaimed relief that it was not a cosmological
disaster~\cite{Weinberg:1983tp}.  
Quickly after that, it
became apparent that not only is the LSP allowed by cosmology, but it
might be preferred since its interaction strength with other particles
can be just right to provide a good cold dark matter 
candidate~\cite{Goldberg:1983nd}. 

I would say that the above story parallels the successful beginning history
of many good ideas.  The LSP was not dreamed up to solve the cold dark
matter problem like other candidate particles were, 
but rather fell out of a self-consistent complete picture
of the supersymmetric theory when attempting to solve its own consistency
issues. And keep in mind, dark matter is a real observational and
experimental problem
that does not seem to be explainable within the standard model.
Nevertheless, it remains to be seen if the LSP is truly the dark matter.  

\section{\bf How can supersymmetry be discovered?}

Post-facto explanations for electroweak symmetry breaking, Higgs sector
quantum stability, Yukawa coupling ratios, and gauge coupling unification
are encouraging, but they are not the end of the story. Additional
phenomena confirming the explanations are required.  
Supersymmetry in fact has many unique and discerning signatures
at high energy colliders, and elsewhere, to solidify its claims on nature.
This is in stark contrast to many speculative theories that
have no correlating phenomena and follow the 
``one observation, one explanation'' pattern.

I will not attempt to go into any detail how supersymmetry
could be found at high-energy colliders.  My approach here is somewhat
paradoxical because more studies and more solid analysis have been plied
to this question than any other, in my estimation. I simply refer the
reader to the many excellent reviews on the subject~\cite{Carena:1999mb}.
These studies show that careful predictions and anticipations
combined with careful experimental preparation and analysis are all needed
to find supersymmetry and accurately interpret new signals. 
Theorists and experimentalists engaged in any of these aspects contribute
substantially to the resolution of the question posed for this section.

We should not forget that many smaller
scale experiments, such as electric dipole moment measurements and
$g-2$ measurements, contribute to our understanding of the natural world,
and by implication the allowed form of a valid supersymmetric theory.
Another important non-collider experiment includes the search for dark matter.
These searches take on many guises, including cryogenic table top
experiments, annihilations into photons in the galactic halo, and
antiproton searches in the galactic halo.  The prediction of many
supersymmetric models is that evidence of LSP scattering would appear
when the experiments get an order of magnitude or two more sensitive.

\section{\bf How can supersymmetry be killed? }

A common concern for any new theory is how it can be killed, or falsified.
This is a complex issue that must combine technological capabilities
(accelerators) with theoretical prejudices (what scale is supersymmetry?).
This question, in my view, has value only inasmuch as it forces the
respondent to formulate more detailed questions on the theory.
For example, an easier subsidiary question would be ``How can we rule
out the bino LSP as an explanation for the dark matter?''
or ``How can we rule out supersymmetry as having an important role
in electroweak symmetry breaking?'' or ``How can we determine if
apparent unification of gauge couplings in supersymmetry is indeed
just an accident?''  These questions are still hard but they are easier
breakdowns of the looming question: how do we determine if all of this
is bunk?

These questions necessarily take on a negative tone, and are not pleasant
to confront for a supersymmetry enthusiast.  An unacceptable answer,
but a technically correct answer, would be ``we will never know.''
As scientists, we must make value judgements on how best to lasso 
natural law and try our best to figure it all out.  
Quixotic pursuits of nonviable theories run counter to these goals.

The community has set several standards on answering these
questions.  One is the careful evaluation of supersymmetric parameter 
space measured against a naturalness or finetuning criteria.  For example,
if supersymmetry
governs electroweak symmetry breaking, the couplings and masses, it is
argued, must be within reasonable bounds.  Numerous papers have
studied these questions in detail~\cite{naturalness}. 
Although they are oftened phrased
positively (``We should see supersymmetry at future collider X''), they
set the tone for when people should give up on
supersymmetry.  Let's call this the ``relevance standard,'' because if
supersymmetry parameters are outside the specified range for relevance to a
problem, it is no longer motivated.

We are currently far from straying outside
the ``relevance standard'', as judged by criteria set forth in the
current literature {\it and} the criteria implied in the original
works on supersymmetry phenomenology~\cite{haber kane}.  If the parameter
space of relevant supersymmetry were mapped as a long drive between
San Francisco and New York City, our current experiments have taken us only
about as far as lowly Elko, Nevada.  

Note also, it is vitally important to the
integrity of the ``relevance standard'' that question {\bf (3)} is answered in
utter completeness, since we continue to find forms of supersymmetry
that are extremely difficult to discover at colliders.  For example, in many
forms of
anomaly mediated supersymmetry breaking, the lighter sparticles are nearly
degenerate, and the lightest sparticle is invisible to the 
detectors~\cite{wino}.
Therefore, all that is produced in high energy collisions at, say, the LHC are
a couple of very soft pions swamped by enormous backgrounds that cannot
be overcome.  Partial progress has been made on 
this issue, but other difficulties have been recognized and there are
surely more discovery subtleties that we have not yet contemplated.

An even stricter application of the ``relevance standard'' can be
found in considering the question, 
``What experiments must be performed to rule out supersymmetric theories 
that produce a bino with cosmologically 
significant cold dark matter density?''  Depending on the precise nature
of the theory, very detailed answers can be given and predictions on
the success of colliders in finding such theories can be 
promulgated~\cite{ellis}.

There is one other standard, which I will call the ``technology standard.''
This is a very practical standard, where 
you might hear it being applied when someone says, ``If the LHC
and NLC do not find supersymmetry I will give up on supersymmetry.'' 
On the surface
it sounds silly, because nature does not care if the accelerator guy
can only squeeze 7 TeV out of the beams.  But without additional
planned accelerators in the lifetime of that physicist, there may not
be need
to address the prospects of supersymmetry manifesting itself at higher
energies. All questions related to supersymmetry become moot to such
a hardline phenomenologist. 
I do not like this standard for many reasons, some of which are
implicit above.  The main argument against it is that each generation
applying it puts an arbitrary
divider line on theory viability based on what they think the last machine,
or last probing experiment can be accomplished in their professional lifetime.
There is room to motivate and encourage better technology and better ideas,
and supersymmetry might be a part of that argument in the future, or it
might not.  But the argument will always need to be made.

\section{Conclusion}

There remain several more questions presented to me at the Institute
meeting that I would like to address in the conclusion.  Many people would
like to know from supersymmetry phenomenologists 
what percentage chance we give supersymmetry of being right.
I have no idea.  Supersymmetry is not
running for city council against other well-defined candidates. 
The only thing I would say is that
it appears to be the most attractive next step in our quest to get a 
deeper understanding of the natural world.  I could never feel confident
in judging its potential success on a more absolute scale because 
I am certain that I have not imagined all 
possibilities.

The next most asked question at the Institute 
along these lines was ``What if supersymmetry
isn't right? Doesn't that mean a lot of people have really wasted their
time?''  
One answer to this question is that
supersymmetry phenomenologists who study how supersymmetry should show up 
at experiments in general, and high energy colliders in particular,
play a major role in helping experiments.  The helpful input spans the
full range from detector design to better techniques for data analysis
and signal extraction.  Remember, these experiments are multi-million
dollar machines, sometimes multi-billion dollar machines, and having
a few theorists around who are actively helping in this regard can
only be a good thing.  It increases the chances of finding a new signal,
whether it be supersymmetry or not, and it helps maximize the interpretative
scope of the data already taken (e.g., are $\mu =0$ theories viable?).
These skills and activities are then transferrable over a wide range of
theory-experiment searches for new physics, whatever the underlying 
theory might be.

A more substantive reply to this question must quarrel with the premise.
The subtle premise of the question is that all worthwhile activity
must be guaranteed at the start to terminate in 
a usable product or textbook theory.  A quick survey
of the entire scientific endeavor, not just particle physics, demonstrates
the folly of this attitude.  For example, if you are attempting to find a
drug that cures a disease, you do not require a guarantee of success
before starting.  You start your work with a thorough knowledge of all
the past sputtered attempts and you make regular evaluations of your work
to determine if your approach can still lead to success.  You get even
more encouraged, and the pharmaceutical company gives you raises, if
it is clear that your approach appears to be not only viable but 
more promising than anyone elses'.  In the end, some of us will find
the drugs that work, some of us won't, and hopefully very few of
us will kill our trial subjects.  But progress is made as a community of
researchers trying to answer well-posed questions from different angles.

Our field is the high-energy frontier, and our questions
are ``What causes electroweak symmetry breaking?'', ``What is the
dark matter?'', ``Why are the neutrinos so light?'', ``What causes CP
violation?'', and 
``What's next on the horizon?'' 
Although many of us work in different aspects
of supersymmetry phenomenology because we
think it is promising, the experiments are what keep us tethered.

We are wed to the questions, not the hypothesized answers.


\end{document}